\documentclass{article}
\usepackage[utf8]{inputenc}
\usepackage{authblk}
\usepackage{setspace}
\usepackage[margin=1.25in]{geometry}
\usepackage{graphicx}
\usepackage{subcaption}
\usepackage{amsmath}
\usepackage{lineno}
\usepackage{multirow}
\usepackage{chngpage}
\usepackage{float}
 \usepackage{adjustbox}
\usepackage{xcolor}
\usepackage{soul}
\usepackage{cite}
\usepackage{comment}
\usepackage{longtable}
\usepackage{braket}
\usepackage[most]{tcolorbox}
\usepackage{amsmath}

\graphicspath{ {./figures/} }

\usepackage[numbers]{natbib}

\title{Role of Quantum Coherence in Chirped Dynamic Nuclear Polarization}

\author[1]{Mayur Jhamnani}
\author[1]{Sajith V Sadasivan}
\author[2]{Sheetal Kumar Jain}
\author[1,3*]{Asif Equbal}

\affil[1]{Center for Quantum and Topological Systems, New York University Abu Dhabi, PO Box 129188, Abu Dhabi, United Arab Emirates.}
\affil[2]{Solid State and Structural Chemistry Unit, Indian Institute of Science, PO Box 560012 Bangalore, India.}
\affil[3]{Department of Chemistry, New York University Abu Dhabi, PO Box 129188, Abu Dhabi, United Arab Emirates.}

\affil[*]{Corresponding author. Email: asif@nyu.edu}
\date{}
\begin{document}
\maketitle
\begin{abstract}
Dynamic Nuclear Polarization (DNP) is transforming NMR and MRI by significantly enhancing sensitivity through the transfer of polarization from electron spins to nuclear spins via microwave irradiation. However, the use of monochromatic continuous-wave (CW) irradiation limits the efficiency of DNP for systems with heterogeneous broad EPR lines. Broad-band techniques such as chirp irradiation offer a solution, particularly for Solid Effect (SE) DNP in such cases. Despite its widespread use, the role of quantum coherence generated during chirp irradiation remains unclear, even though it is a key factor in determining the maximum achievable DNP efficiency. In this work, we use density matrix formalism to provide a comprehensive understanding of the quantum coherence generated during electron-nucleus double-quantum (DQ) and zero-quantum (ZQ) SE transitions and their impact on Integrated Solid Effect (ISE) DNP under chirp irradiation. Our analysis employs fictitious product-operator bases to trace the evolution of electron-nucleus coherence leading to integrated or differentiated SE. We also explore the previously unexamined role of decoherence in optimizing chirped DNP. These findings provide new insights into low-temperature DNP and triplet DNP using photoexcited electrons.

\end{abstract}

\linespread{1}
\section{Introduction} \label{sec:Introduction}

\begin{figure}[h!]
    \centering
    \includegraphics[width = 0.9\textwidth]{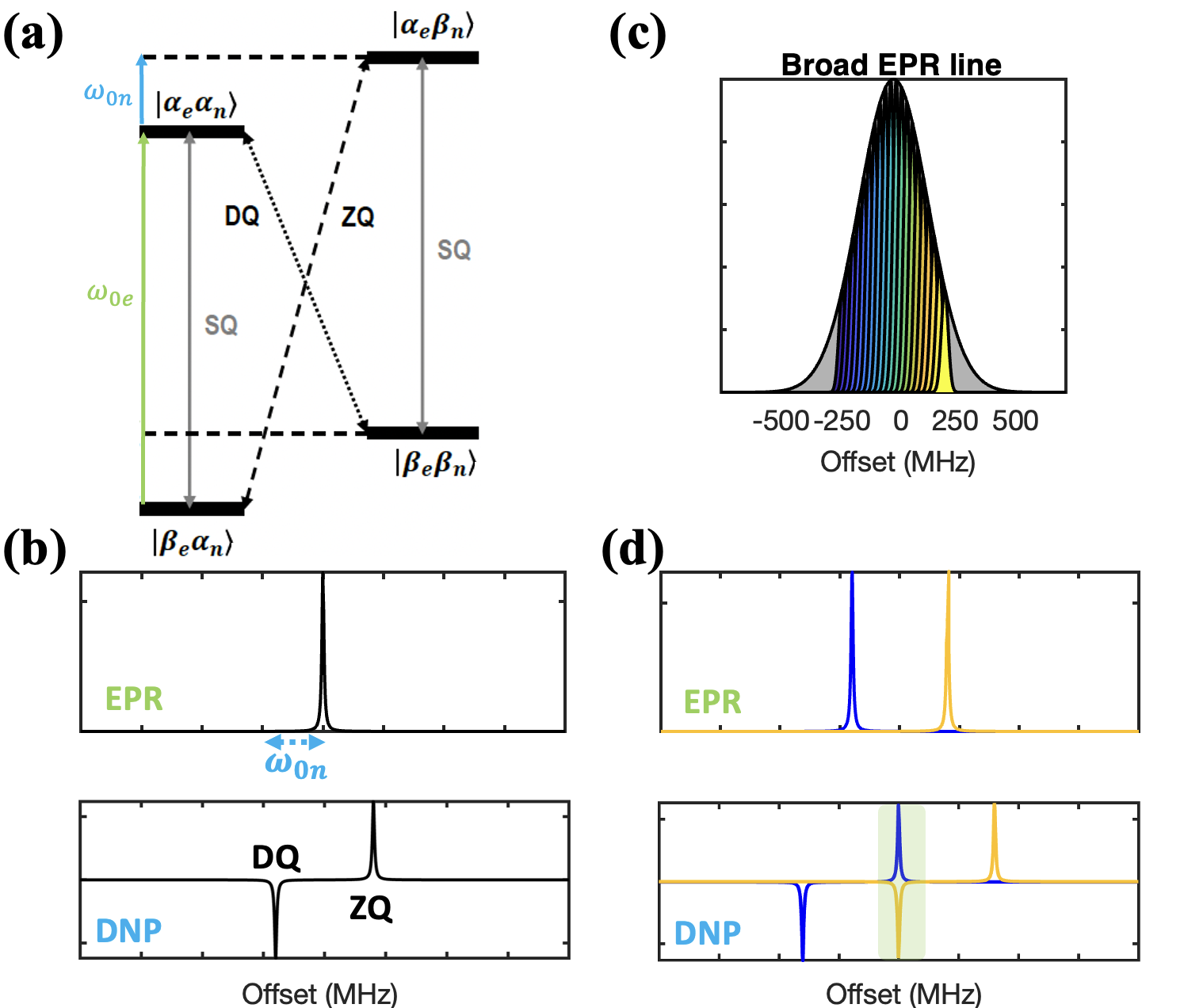}
    \caption{(a) A schematic diagram of the energy levels and transitions in a coupled $e-n$ system, each having spin 1/2. (b) The SE DNP enhancement profile with respect to offset illustrating the polarization enhancement from the DQ and ZQ transitions for a single electron spin packet coupled to a nuclear spin. The respective EPR spectrum is shown above for comparison.  (c) A schematic representation of the individual electron spin packets that contribute to the inhomogeneously broadened EPR line is shown. (d) The concept of DSE is illustrated based on the overlap of DQ and ZQ transitions of different spin packets leading to the cancellation or reduction of SE DNP enhancement at the highlighted offset.}
    \label{fig:introduction_fig}
\end{figure}

Nuclear Magnetic Resonance (NMR) spectroscopy is an indispensable noninvasive tool that provides detailed insights into both the structure and dynamics of a wide range of samples.\cite{ernst1990principles,kay2011nmr,hashemi2012mri,qiang2017structural,grey2004nmr} However, the technique suffers from an inherent low signal sensitivity as a result of the low thermal polarization of nuclear spins.\cite{abragam1961principles} To address this issue, nuclear spin hyperpolarization techniques are utilized.\cite{abragam1978principles,eills2023spin,maly2008dynamic,barnes2008high} Dynamic nuclear polarization (DNP) is one of the most widely used hyperpolarization methods that transfer polarization from the unpaired electron spins of a paramagnetic center to nuclear spins through microwave ($\mu w$) irradiation.\cite{carver1953polarization,overhauser1953polarization} The larger gyromagnetic ratio of electron spins results in a higher Boltzmann polarization, which can be transferred to nuclear spins via DNP.\cite{overhauser1953polarization,hovav2010theoretical,hovav2012theoretical,thankamony2017dynamic} Among the various DNP mechanisms, the Solid Effect (SE) is considered the simplest in insulating solids.\cite{hovav2010theoretical,hu2011quantum,corzilius2012solid} SE DNP involves a dipolar-coupled electron and nuclear ($e$-$n$) spin system, where an off-resonant $\mu w$ irradiation with an offset of $\omega_{0n}$ with respect to the electron Larmor frequency ($\omega_{0e}$) induces double-quantum (DQ) and zero-quantum (ZQ) transitions, as shown in Figure \ref{fig:introduction_fig}a. The DQ and ZQ transitions contribute to positive and negative polarization enhancements, respectively (Figure \ref{fig:introduction_fig}b). 

In the case of paramagnetic centers with broad electron paramagnetic resonance (EPR) lines resulting from various anisotropic interactions such as g-tensors and hyperfine coupling, achieving an efficient SE DNP is more complicated. The broad EPR line can be considered to be made up of different spin packets (Figure \ref{fig:introduction_fig}c). Taking into account the influence of all spin packets within the EPR line leads to an overlap of DQ and ZQ transitions from different spin packets. This effect reduces the overall enhancement of SE DNP and leads to self-cancellation of the enhancement (Figure \ref{fig:introduction_fig}d), a phenomenon known as differential SE (DSE). \cite{takeda2001dynamic, hamachi2024recent, wenckebach2008solid} Moreover, the ZQ and DQ SE transitions are quantum mechanically forbidden with respect to $\mu w$ irradiation. Hence, exciting a wide band of forbidden transitions requires significantly higher power levels, which results in additional challenges such as sophisticated $\mu w$ power amplifiers and severe sample heating.

In this context, a more effective approach is to use chirp $\mu w$ irradiation, where the $\mu w$ frequency varies over time, termed chirped DNP.\cite{kaminker2018amplification,equbal2019pulse} Chirped DNP can be achieved using an arbitrary waveform generator (AWG) at a fixed magnetic field, or by sweeping the external magnetic field while keeping the $\mu w$ frequency fixed.\cite{can2018frequency,quan2022integrated} Accordingly, it fulfills two primary objectives. Firstly, it enables broadband excitation with minimal $\mu w$ power, inheriting the contributions from a large number of electron spin packets across a broad EPR spectrum. Secondly, a frequency sweep across the resonance frequencies, fulfilling the ZQ and DQ resonance conditions for each spin packet, can potentially integrate the ZQ and DQ enhancements of each individual spin packet. 

Henstra \textit{et al}. introduced an innovative method known as the integrated solid effect (ISE), designed specifically to enhance DNP in paramagnetic centers that exhibit an inhomogeneous broad EPR line, such as photoexcited triplet states.\cite{henstra1988enhanced} The process entails a linear sweep of the magnetic field from a large negative to a large positive offset, or the opposite. The sweep encompasses the DQ-SE condition, the EPR on-resonance SQ condition, and the ZQ-SE condition across all spin packets present in the EPR spectrum. The magnetic field (or $\mu w$ frequency) sweep results in three events: the DQ transition, the adiabatic inversion of the electron spin polarization, and the ZQ transition. The adiabatic inversion of the electron spin causes the following ZQ enhancement to integrate in the same direction as the DQ enhancement, thereby presumably integrating the two SE DNP enhancement and combining the contributions from all electron spin packets, leading to substantial DNP enhancement per unit time.
\cite{henstra1988enhanced, HENSTRA19906, henstra2014dynamic} Experimentally, ISE is utilized for DNP transfer in a photoexcited triplet system where orders of higher enhancement can be obtained using non-thermally polarized electron spins. Henstra \textit{et al.} obtained a 5,500-fold enhancement of $^1H$ polarization, as compared to the theoretical limit of 660 in the conventional DNP techniques.\cite{HENSTRA19906} Takeda \textit{et al}. obtained an $^1H$ enhancement of 3,160 for using pentacene-doped naphthalene at 100 K in a field of 0.3 T.\cite{takeda2001dynamic} In another study, Iinuma \textit{et al}. obtained a record $^1H$ DNP enhancement of 80,000 at 77 K and 0.3 T. \cite{iinuma2000high} 
Takeda \textit{et al.} also offered an interesting perspective on ISE and introduced an alternative term, "integrated cross-polarization" (ICP), instead of ISE, due to its strong similarity to the cross-polarization (CP) methodology used in NMR.\cite{takeda2001dynamic} Recently, the ISE DNP method has been used for biomedical quantum sensing applications through the development of tailored materials that host organic chromophores and generate a triplet electron spin system upon photoexcitation.\cite{nishimura2020materials,hamachi2024recent} 

In addition, Griffin and co-workers have extensively utilized chirp $\mu w$ irradiation for pulsed DNP experiments.\cite{can2018frequency,tan2020adiabatic,quan2022integrated,quan2023general} Two new schemes, named adiabatic SE (ASE) and stretched SE (SSE) have been recently introduced. ASE employs a narrow sweep that targets a specific SE matching condition (ZQ or DQ), while SSE uses a broad sweep through one of the SE matching conditions.\cite{can2018frequency, tan2020adiabatic, quan2022integrated} 

Although Chirped DNP has led to better polarization transfer, there is still room for improvement in the understanding of its working mechanism. Specifically, the influence of the quantum coherence generated during the three events on the maximizing of DNP transfer needs further investigation. Currently, the understanding of chirped DNP is based exclusively on the polarization of the electron spin.\cite{hautle2024creating} The role of coherence generated during chirp has been ignored so far. Quantum coherence is also fundamental in quantum information science for stabilizing and manipulating quantum states with high precision, essential for emerging quantum technologies.

In our study, we observe that the occurrence of ISE is not consistent with the parameters of chirp irradiation, regardless of the adiabatic inversion of electron spin polarization. The theoretical explanation based only on spin polarization does not account for why chirped DNP occasionally results in integrated or, at other times, in differentiated DQ and ZQ enhancements within the coherent chirped DNP regime. As a result, we have used a simplified spin-1/2 $e$-$n$ spin system to explain the crucial role of various $e$-$n$ coherences in the spin dynamics during the chirp.

In the next sections, we first establish the resonance conditions for ZQ and DQ SE. Subsequently, we examine the impact of chirp on a quantum state of the system or density matrix that includes polarization and coherence. We explore the influence of both adiabatic and non-adiabatic sweeps on the density matrix for both SQ and DQ/ZQ transitions. We utilize the reduced-density matrix formalism to elucidate three fundamental processes within ISE DNP. We hitherto elucidate how decoherence can maximize chirped DNP enhancement, providing a new perspective for the progress of triplet-DNP. Our theoretical examination also sheds light on ASE DNP.

\section{Theory and Mechanism} \label{sec:Theory and Mechanism}

To understand the spin dynamics during Chirp-DNP, we employ a simplified spin-1/2 system composed of dipolar-coupled electron ($\mathbf{S}$) and nuclear ($\mathbf{I}$) spins ($e-n$). We begin by examining the most general form of a coupled $e$-$n$ spin system under continuous-wave (CW) $\mu w$ irradiation conditions and derive the effective Hamiltonian leading to ZQ and DQ SE. 

\subsection{Solid Effect}
\label{sec:The Solid Effect}

We consider the general form of Hamiltonian governing the SE for an $e-n$ system in the laboratory frame of nuclear spins and rotating frame of electron spins, under $\mu w$ irradiation.
\begin{equation}\label{se_H}
    H_0 =  (\omega_{0e} -\omega_{\mu w}) S_z  +\omega_{0n}I_z\\ + AS_zI_z + BS_zI_x + \omega_1 S_x
\end{equation}
Here, $(\omega_{0e} - \omega_{\mu w})$ represents the offset with respect to the $\mu w$ frequency for the electron spins whereas $\omega_{0n}$ and $\omega_1$ denote the nuclear Larmor frequency and the power (amplitude) of the applied $\mu w$ respectively. The coefficients $A$ and $B$ represent the secular and pseudo-secular hyperfine coupling (HFC) terms, respectively, between electron and nuclear spins. The term $B$ is anisotropic and can be described by angle $\beta$ with respect to the external magnetic field. The pseudo-secular term does not commute with the nuclear Zeeman interaction. Consequently, it is crucial to perform the analysis in the nuclear laboratory frame. The Hamiltonian can be simplified by a series of unitary transformations and an effective Hamiltonian can be derived using average Hamiltonian theory. The effective Hamiltonian can be analysed using DQ and ZQ fictitious spin-1/2 operators given by $DQ_X = S_xI_x - S_yI_y$, $DQ_Y = S_xI_y + S_yI_x$, $DQ_Z = \frac{1}{2}(S_z + I_z)$) and  $ZQ_X = S_xI_x + S_yI_y$, $ZQ_Y = S_xI_y - S_yI_x$, $ZQ_Z = \frac{1}{2}(S_z - I_z)$ respectively. The effective SE DNP Hamiltonian can be expressed as:
\begin{align}
        \Tilde{H}_{ZQ-SE} &= \frac{\omega_1 B}{4\omega_{0n}}(S^+ I^- + S^- I^+)=\frac{\omega_1 B}{2\omega_{0n}}(S_x I_x + S_y I_y) = \frac{\omega_1 B}{2\omega_{0n}}ZQ_x\\
        \Tilde{H}_{DQ-SE} &= \frac{\omega_1 B}{4\omega_{0n}}(S^+ I^+ + S^- I^-) =\frac{\omega_1 B}{2\omega_{0n}}(S_x I_x - S_y I_y)=\frac{\omega_1 B}{2\omega_{0n}}DQ_x
    \label{subspace_Hamiltonian}
\end{align}

The effective Hamiltonian remains time independent when ${\sqrt{ {(\omega_{0e} - \omega_{\mu w})}^2 + {\omega_{1}}^2 } =\omega_{0n}}$, subjected to DQ or ZQ matching conditions. We can calculate nuclear spin polarization, by evolving the initial density matrix $\rho (0) = {S_z}$ under the effective SE DNP Hamiltonian and finding the projection of the time-dependent density matrix on to $I_z$ (nuclear spin polarization operator).

\begin{align}
    \nonumber \left\langle {\hat I_z} \right\rangle  &= Tr[\hat I_z\rho (t)]\\
\text{where, }\rho (t)&= {e^{ - i{\Tilde{H}_{SE}}t}}\rho (0){e^{i{\Tilde{H}_{SE}}t}}
\end{align}

Using $\rho (0) = {S_z}$ = $({\textstyle{{{S_z} + {I_z}} \over 2}}) + ({\textstyle{{{S_z} - {I_z}} \over 2}}) = DQ_z + ZQ_z$, the evolution of density matrix ($\rho$) can be derived from the Liouville von-Neumann equation. Since the ZQ and DQ subspaces commute, $\rho$ can also be divided into two subspaces similar to the Hamiltonian.\cite{hu2011quantum,corzilius2012solid,jain2017off,pang2022unified} Notably, each subspace behaves like an effective spin-1/2 system, where the corresponding fictitious Pauli operators are used to represent the dynamics within that subspace.\cite{vega1978fictitious}

\begin{align}
\nonumber {\tilde\rho} (t)_{DNP}^{ZQ} =&   ({\textstyle{{{S_z} + {I_z}} \over 2}}) + ({\textstyle{{{S_z} - {I_z}} \over 2}})\cos ({\textstyle{{{\omega_1 B_{}}t} \over 2 \omega_{0n}}}) - ({\textstyle{{S_{}^ + {I^ - } - S_{}^ - {I^ + }} \over {2i}}})\sin ({\textstyle{{{ \omega_1 B_{}}t} \over 2 \omega_{0n}}})\\
=&   DQ_z + ZQ_z \cos ({\textstyle{{{\omega_1 B_{}}t} \over 2 \omega_{0n}}}) - ZQ_y \sin ({\textstyle{{{ \omega_1 B_{}}t} \over 2 \omega_{0n}}})\\
\nonumber {\tilde\rho} (t)_{DNP}^{DQ} =&   ({\textstyle{{{S_z} - {I_z}} \over 2}}) + ({\textstyle{{{S_z} + {I_z}} \over 2 }})\cos ({\textstyle{{{\omega_1 B_{}}t} \over 2 \omega_{0n} }}) - ({\textstyle{{S_{}^ + {I^ + } - S_{}^ - {I^ - }} \over {2i}}})\sin ({\textstyle{{{\omega_1 B_{}}t} \over 2\omega_{0n}}})\\
  =& ZQ_z + DQ_z\cos ({\textstyle{{{\omega_1 B_{}}t} \over 2 \omega_{0n}  }}) - DQ_y\sin ({\textstyle{{{\omega_1 B_{}}t} \over 2\omega_{0n}}})
\end{align}

Clearly, when ${\textstyle{{{\omega_1 B_{}}t} \over 2\omega_{0n}}t} = \pi$, ${\tilde\rho} (t)_{DNP}^{ZQ} $ = ${I_z}$, implying the rotation of $ZQ_z$ along $ZQ_x$ by 180$^{\circ}$. This transfers electron spin polarization ($S_z$) to nuclear spin polarization ($I_z$). Similarly electron spin polarization is converted to negative polarization of nucleus -${I_z}$ by 180$^{\circ}$ rotation of $DQ_z$ along $DQ_x$. A rotation $\neq n.\pi$ generates coherence between the electron and nuclear spin. The rotation around the $ZQ_x$ or $DQ_x$ axis proceeds at a frequency given by ${\textstyle{{{\omega_1 B_{}}t} \over 2\omega_{0n}}}$, which is quite small when $\omega_1$ is significantly less than $\omega_{0n}$, a common case in high magnetic field DNP. The consequences of this will be explored in the next subsection, where we demonstrate how the polarization and coherence of the density matrix evolve with the applied chirp pulse swept across the Larmor frequency of a single electron spin. This analysis will help in evaluating analogous situations within the $e$-$n$ DQ and ZQ subspaces in Section \ref{sec:results}.

\begin{figure}[h!]
    \centering
    \includegraphics[width=\textwidth]{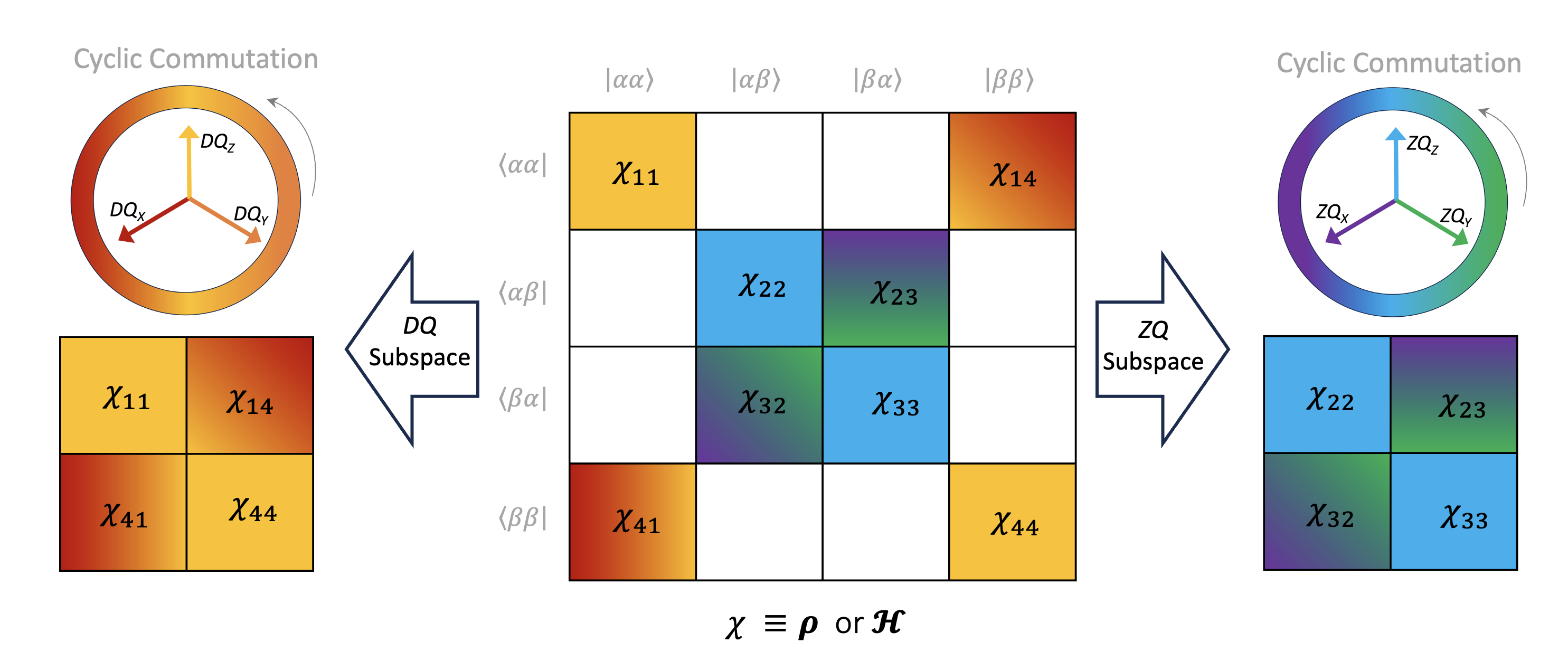}
    \caption{Matrix representation of the full density matrix, along with the reduced ZQ and DQ density matrices. The diagonal entries indicate populations, while the off-diagonal elements represent coherences.}
    \label{matrix_representation}
\end{figure}

\subsection{Adiabatic and Non-Adiabatic Transitions}\label{sec:Adiabatic Inversion}

Adiabatic sweeps using chirped pulses are a powerful technique in magnetic resonance in scenarios where there is inhomogeneous broadening in the spin system, such as variations in the g-tensor. In this approach, the frequency of the applied pulse is varied smoothly over time, typically in a linear fashion. This frequency sweep allows for the adiabatic passage of spins through resonance, resulting in population inversion or other desired state transformations depending on the sweep pulse parameters. Here, we understand the parameters determining the adiabatic nature of a linear chirp pulse on a general initial density matrix.

When a single electron spin \textbf{S} is subjected to a $\mu w$ chirp pulse in an external magnetic field, the Hamiltonian in the electron spin rotating frame is given by: 
\begin{equation}
    \label{chirp_Hamiltonian}
	H_{chirp} = (\omega_{0e} -\omega_{\mu w} (t)) S_z + \omega_1 S_x
\end{equation}
where the coefficients ($\omega_{0e} -\omega_{\mu w} (t) =\Omega_{e}(t)$) represents the time-dependent offset and $\omega_1$ represents the $\mu w$ amplitude which acts as a perturbation ($p$) and mixes the eigen-states of Zeeman Hamiltonian. The $\mu w$ chirp pulse can result in an adiabatic or non-adiabatic transition depending on the Landau-Zener (LZ) factor, $\frac{\pi p^2}{2 k}$, where k represents the sweep rate through the resonance condition ($\omega_{0e} =\omega_{\mu w}$). If the LZ factor is much higher than unity, there will be complete inversion of the population difference or the polarization between states. In such a scenario, the transition can be regarded as fully adiabatic. If the LZ factor is small, the transition leads to an incomplete inversion and the generation of superposition states. Such transitions are non-adiabatic. Furthermore, beyond adiabaticity, the initial density matrix, whether representing polarization or coherence prior to the transition, also plays a crucial role in determining the final outcome of the transition induced by the $\mu w$ chirp pulse.

For better visualization of the evolution of the density matrix, it is depicted as a magnetization vector on the Bloch sphere. The $x$, $y$, and $z$ components of the magnetization vector are the expectation values of the Pauli operators ($S_x, S_y, S_z$) in the spin-1/2 basis, or corresponding operators in the reduced fictitious basis. A magnetization vector oriented towards the poles of the Bloch sphere represents a polarization, whereas one oriented towards the equator denotes a coherence. Any orientation between these extremes depicts a mixture of polarization and coherence.

\begin{figure}[h!]
    \centering
    \includegraphics[width = \textwidth]{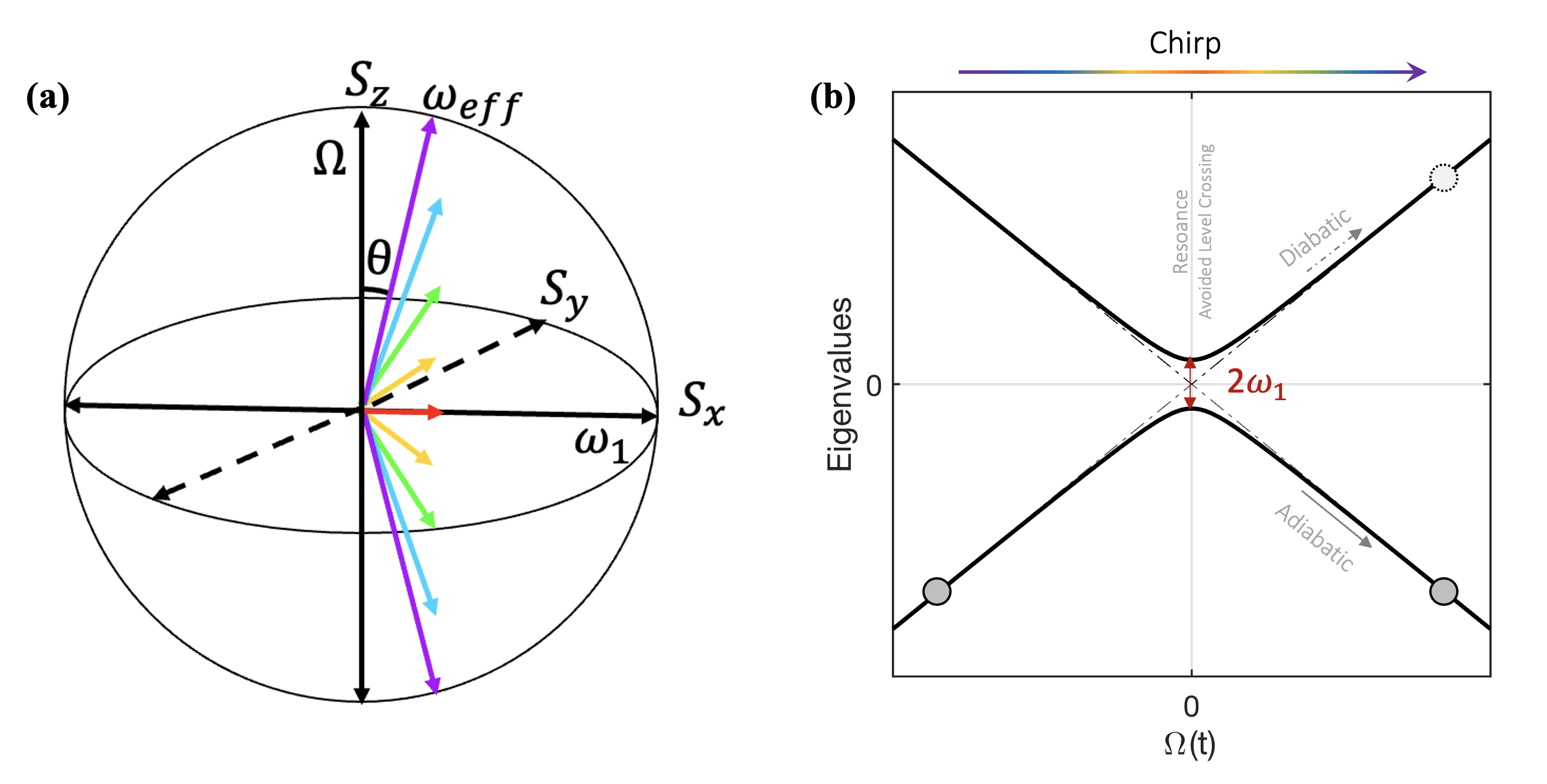}
    \caption{Pictorial representation of (A) the effective frequency vector, and correspondingly, the Hamiltonian in Equation \ref{eff_Hamiltonian}, on the Bloch sphere (B) Time-dependent energy levels of the system, depicting energy level crossing and anti-crossing.}
    \label{fig:bloch_hamiltonian}
\end{figure}

Within the Bloch sphere framework, the chirp pulse is depicted via a time-dependent effective field: $\omega_{eff}(t)$ =$\sqrt {\Omega_e(t)^2 + \omega_1^2}$ where both the magnitude and the direction of the effective field changes. The direction of the effective field rotates, causing the eigen basis of the effective Hamiltonian to rotate during chirp. \cite{ivanov2021floquet} The angle, $\theta$ between $\omega_{eff}(t)$ and $\Omega_e(t)$ is given by $\sin^{-1}(\frac{\omega_1}{\omega_{eff}(t)})$. The Hamiltonian in Equation \ref{chirp_Hamiltonian} can be rewritten as:

\begin{equation}\label{eff_Hamiltonian}
H_{chirp} = \omega_{eff}(t)\cos(\theta(t)) S_z + \omega_{eff}(t)\sin(\theta(t)) S_x
\end{equation}

and this can be geometrically represented on a sphere as shown in Figure \ref{fig:bloch_hamiltonian}a. Clearly, at on resonance conditions, effective field is $\omega_1$ pointing towards $x$ according to Equation \ref{eff_Hamiltonian} and the eigen values are $\pm$$\omega_1$ as depicted in Figure \ref{fig:bloch_hamiltonian}b.

In the following subsections, the evolution of the density matrix under the Liouville-von Neumann equation is examined to gain further insight into the effect of a $\mu w$ chirp pulse on spin \textbf{S} in two cases: when the initial density matrix is (i) purely polarization and when it is (ii) purely coherence.

\subsubsection{Case I: $\mathbf{\rho_0 = S_z}$ (polarization)}

When $\mathbf{\rho_0 = S_z}$, the initial magnetization vector is oriented in the $+z$ direction commuting with Zeeman Hamiltonian. The $\mu w$ chirp pulse rotates this magnetization vector. The trajectory of the magnetization vector depends on the adiabaticity of the pulse and differs between adiabatic and non-adiabatic conditions, which will be elaborated upon below.

\begin{figure}[h]
    \centering
    \includegraphics[width = 0.9\textwidth]{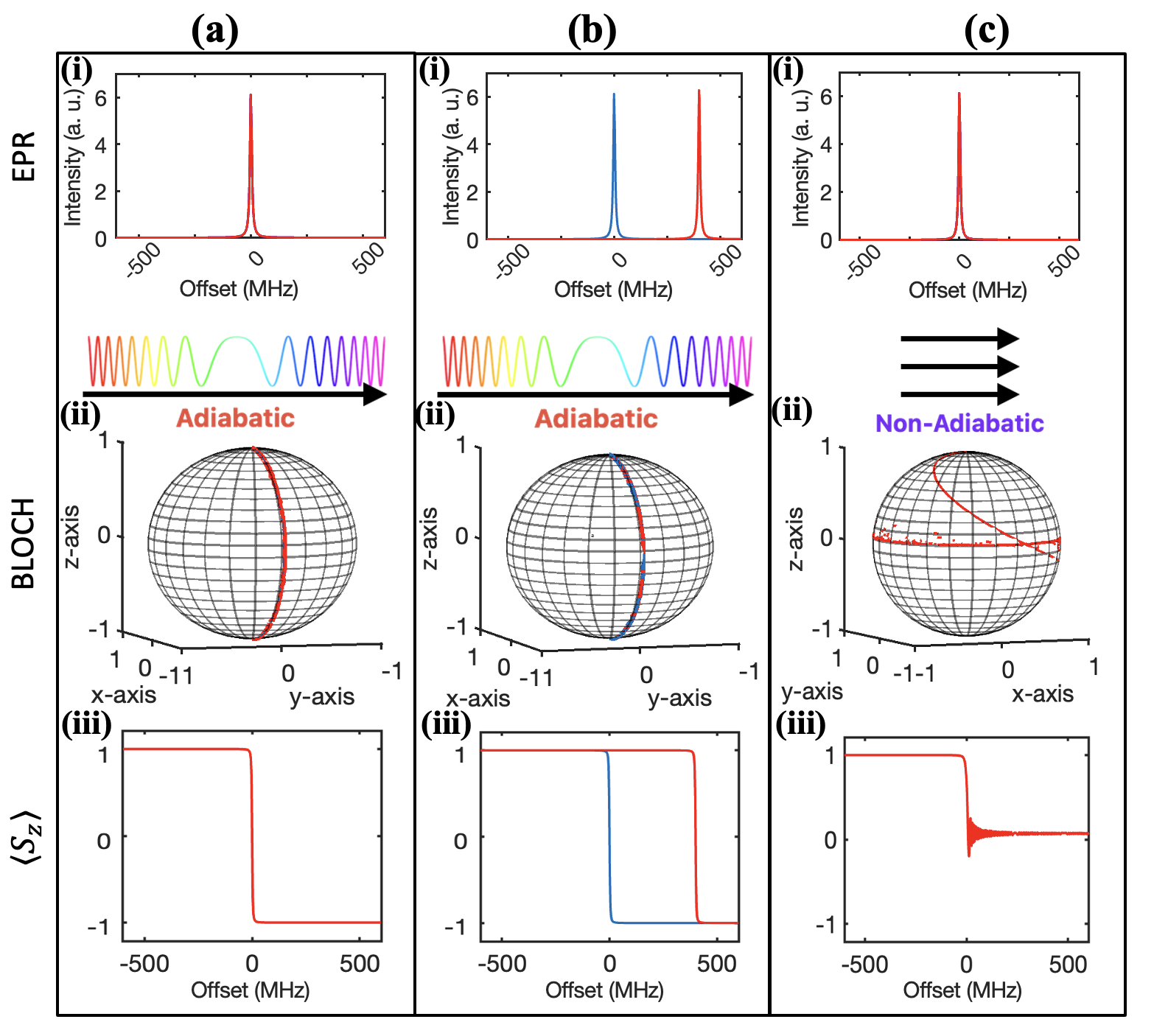}
    \caption{Adiabatic inversion is demonstrated using a $\mu w$ chirp pulse having $\mu w$ power $\omega_1$ = 4 MHz for the case of (a) a single electron spin and (b) two electron spins with Larmor frequencies differing by 400 MHz. The offset is being swept from -600 MHz to 600 MHz at a slow rate of 1.5 MHz$/\mu s$. (c) Incomplete inversion is shown for a scenario where a single electron spin does not meet the adiabatic condition. The offset is being swept at a fast rate of 250 MHz$/ \mu s$.}
    \label{fig:Adiabatic}
\end{figure}

\textbf{a. Adiabatic $\mu w$ Chirp: } For a single electron spin packet, as shown in Figure \ref{fig:Adiabatic}a (i), if the $\mu w$ Chirp is adiabatic, the magnetization vector follows the effective field. It precesses around the effective field vector while the field vector is continuously tipping down (Figure \ref{fig:Adiabatic} a (ii)). The evolution (precession around the effective field and continuous tipping down) occurs due to the small but non-zero commutator between the associated density matrix and the effective Hamiltonian.  As a result, the evolution of the magnetization vector follows the time-dependent effective Hamiltonian. This causes the complete rotation of the magnetization vector from +z to -z. The net coherence is zero at the end of the sweep as seen in Figure \ref{fig:Adiabatic}a(ii-iii).

The real advantage of using an adiabatic $\mu w$ chirp pulse lies in its ability to invert multiple spin packets that resonate at different Larmor frequencies through a single channel irradiation as depicted in Figure \ref{fig:Adiabatic}b(i-iii). This is particularly useful for exciting a large number of spin packets of a paramagnetic center with broad-EPR line. 
\cite{abragam1961principles,baum1985broadband,garwood2001return}

\textbf{b. Non-adiabatic $\mu w$ Chirp: }Given that the initial magnetization vector points in the $+z$ direction, a non-adiabatic $\mu w$ chirp irradiation does not lead to a complete inversion of the magnetization vector. Rather, it causes the final magnetization vector to point in between poles, as shown in Figure \ref{fig:Adiabatic}c (ii), creating a quantum state with both polarization and coherence (Figure \ref{fig:Adiabatic}c(iii)). This happens when the commutator $[H_{eff}, \rho]$ is large and the evolution of the magnetization vector does not follow the time-dependent effective Hamiltonian.

\subsubsection{Case 2: $\mathbf{\rho_0 = sin(\phi) S_x + cos(\phi) S_y}$ (coherence)}

When the initial density matrix is a pure coherence, the magnetization vector points in the equatorial region of the Bloch sphere and is perpendicular to the effective field. As in case I, the trajectory followed by the magnetization vector is influenced by the adiabaticity of the pulse and varies under adiabatic versus non-adiabatic conditions, which will be discussed further below.

\begin{figure}[h!]
    \centering
    \includegraphics[width = 0.9\textwidth]{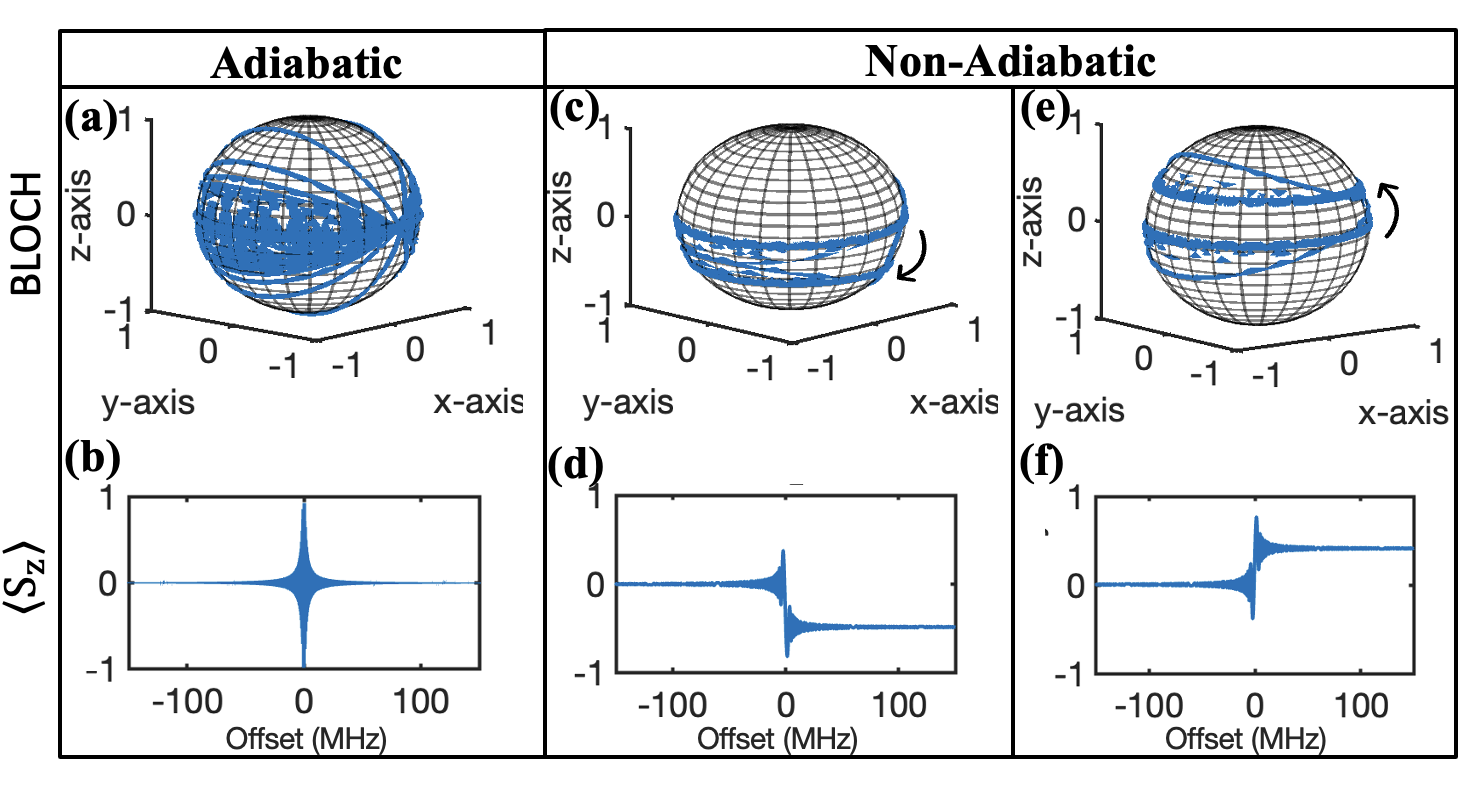}
    \caption{The time evolution of coherences in the density matrix with $\rho_0 = sin(\phi)S_x + cos(\phi)S_y$ upon the application of a $\mu w$ chirp pulse is depicted with (a, b) rate = $1 MHz/\mu s$ and $\phi = 45^o$, (c, d) rate = $12 MHz/\mu s$ and $\phi = 45^o$, and (e, f) rate = $12 MHz/\mu s$ and $\phi = 220^o$. The chirp in (a, b) is adiabatic, satisfying the Landau-Zener condition, while that in (c, d, e, f) is non-adiabatic.}
    \label{fig:adia_non_adia_Sz}
\end{figure}

\textbf{a. Adiabatic $\mu w$ Chirp: } When an adiabatic $\mu w$ chirp pulse is applied, the magnetization vector rotates in a plane perpendicular to the effective field equatorial region. This is because the initial density matrix and the effective Hamiltonian are orthogonal. Moreover, the plane follows the rotation of the effective field - completing one full rotation. 

As a result, there is no net (or negligible) component of the magnetization vector along the z-axis. Therefore, no net polarization is generated when the effective field rotates by 180$^{\circ}$. This can be visualized through the expectation value of $S_z$ plotted in Figure \ref{fig:adia_non_adia_Sz}b. 

\textbf{b. Non-adiabatic $\mu w$ Chirp: }Contrary to the previous case, the outcome of irradiating a non-adiabatic $\mu w$ chirp pulse on a coherence is a magnetization vector with a component along the z-axis, representing a net polarization (Figure \ref{fig:adia_non_adia_Sz}(c, e)). This occurs because the plane does not follow the rotation of the effective field due to the non-adiabaticity of the transition. Moreover, as seen in Figure \ref{fig:adia_non_adia_Sz}(d, f), the net polarization can either be positive or negative depending on the phase, $\phi$ of the coherence just before the transition occurs. 

Overall, we have seen that the outcome of any transition depends on (i) the initial density matrix and (ii) the adiabaticity of the perturbation. This has been summarized in Table 1.

\begin{table}[h!]
    \centering
    \caption{Tabular representation of the outcome of a transition caused by a $\mu w$ chirp pulse. The outcome depends upon the initial density matrix and the adiabaticity of the transition.}
    \begin{tabular}{|c|c|c|}
        \hline
        \textbf{Transition} & \textbf{Inital Density Matrix} & \textbf{Final Density Matrix} \\
        \hline
        Adiabatic & Polarization & Polarization \\
        \cline{2-3}
                  & Coherence & Coherence \\
        \hline
        Non-Adiabatic & Polarization & Polarization + Coherence \\
        \cline{2-3}
                      & Coherence & Polarization + Coherence \\
        \hline
    \end{tabular}
\end{table}

In the forthcoming section, this concept is applied to DNP transitions (DQ and ZQ) caused by an effective Hamiltonian which depends on a $\mu w$ chirp pulse (Chirped-DNP) as well as the hyperfine coupling as seen in Equation 2-3.

\section{Results and Discussion}\label{sec:results}

This section extends our discussion by illustrating the significance of the quantum coherences that are generated when a dipolar coupled $e-n$ system (\textbf{S}=1/2, \textbf{I}=1/2) is subjected to a $\mu w$ chirp pulse. Here, we explore ISE, which is an experimentally utilized broadband chirped DNP scheme, to understand how the adiabaticity of $\mu w$ chirp pulse inducing different transitions impacts the net DNP transfer. In ISE, the offset is swept through across both DQ and ZQ transitions. The Hamiltonian governing  chirped DNP can be framed by introducing the Zeeman interaction of nuclear spin having Larmor frequency $\omega_{0n}$ and hyperfine coupling terms, $AS_zI_z + BS_zI_x$, as shown in Equation \ref{chirp_DNP_Hamiltonian}:
\begin{equation}\label{chirp_DNP_Hamiltonian}
      \Tilde{H}_{ISE}(t) = \omega_{eff} (t) cos(\theta(t))S_z + \omega_{eff} (t) sin(\theta(t)) S_x  - \omega_{0n}I_z + AS_zI_z + BS_zI_x.
\end{equation}
Assuming that the spin system is thermalized with the internal Hamiltonian, the initial density matrix is $\rho(0) \approx S_z$. This is valid under the high-field approximation.

\begin{figure}[h]
    \centering
    \includegraphics[width = 0.5\textwidth]{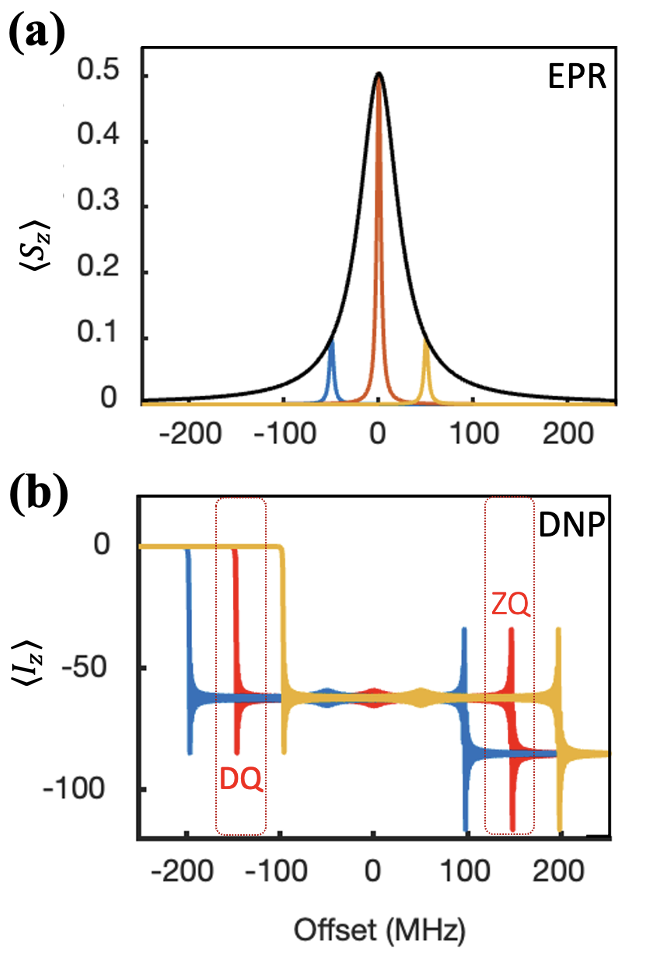}
    \caption{Illustration of ISE of a broad EPR spectrum: (a) The contribution from individual spin packets, and (b) the DNP profile in the ISE regime is depicted. $\omega_{0n} = 146.8 MHz$.sweep range $-300$ to $300 MHz$. $\omega_1 = 8MHz$. $k = 1.5MHz/\mu s$.}
    \label{fig:ISE_broad_EPR}
\end{figure}

\begin{figure}[h!]
    \centering
    \includegraphics[width = \textwidth]{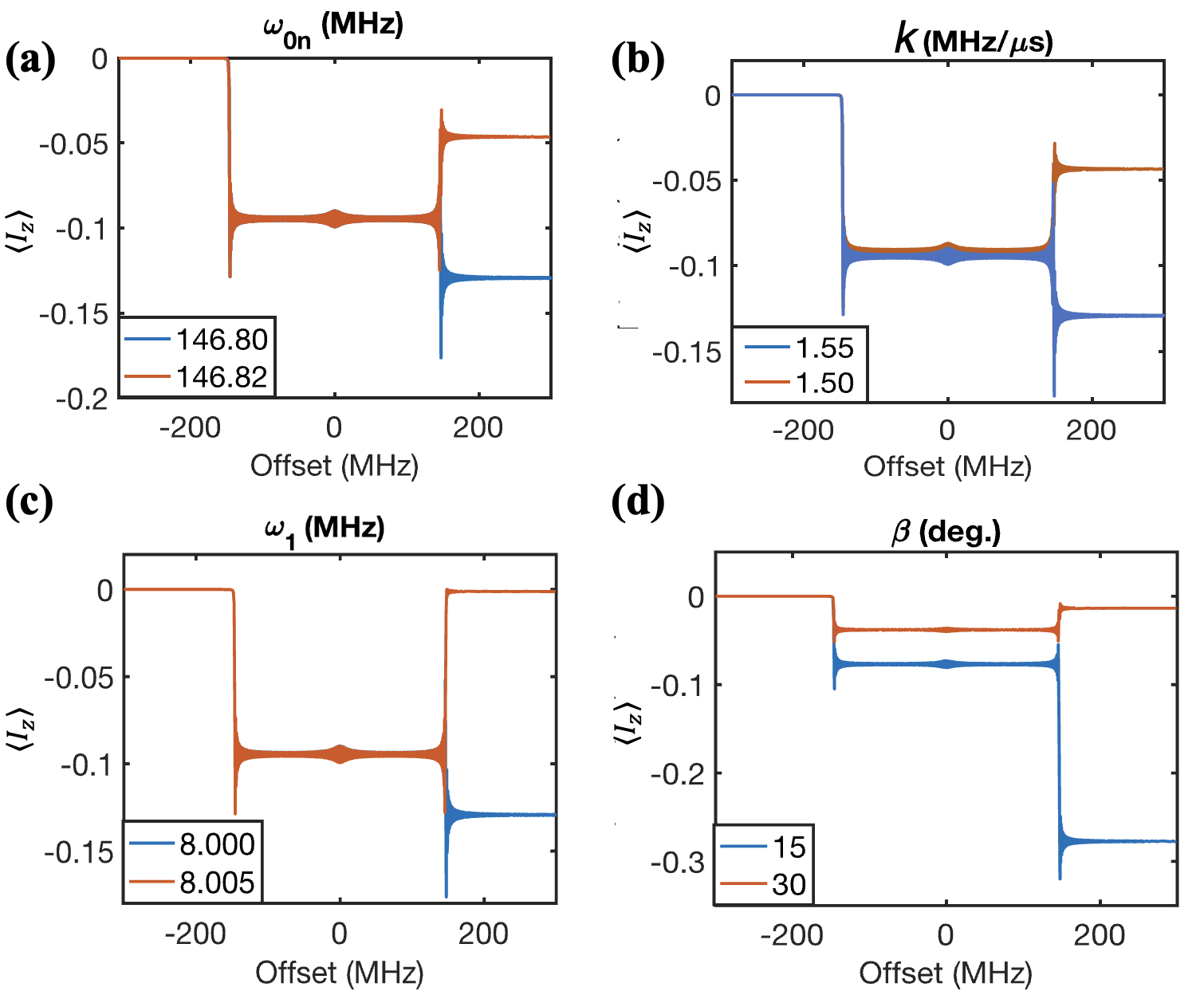}
    \caption{Depiction of ISE (blue) and DSE (orange). Other than electron Larmor frequency, even a slight change in experimental parameters such as (a) nuclear Larmor frequency, $\omega_{0n}$, (b) Sweep rate, k, (c) $\mu w$ amplitude, $\omega_1$, or (d) e-n orientation, $\beta$ impacts the direction of the ZQ transition, and affects ISE. Sweep range: $-300 MHz$ to $300 MHz$.}
    \label{fig:ISE_other_factors}
\end{figure}

Upon numerically simulating ISE for a broad EPR line paramagnetic center, we observed that the $\mu w$ chirp pulse can potentially cause the integration of the DQ enhancement with the ZQ enhancement for all the spin-packets spanned by the width of the chirp. This is shown in Figure \ref{fig:ISE_broad_EPR}(a, b) where both DQ and ZQ polarization transfer occur in the same direction for all the spin packets. We expected this to occur consistently for different experimental conditions. However, when experimental parameters determining LZ factors (SE perturbation and chirp sweep rate) such as $\omega_{0n}$, $\mathit{k}$, $\omega_1$, and $\beta$ were slightly varied, we observed a caveat. The polarization initially transferred from \textbf{S} to \textbf{I} during the DQ transition sometimes reverted back to \textbf{S} during the ZQ transition depending on the experimental parameters. In such scenarios, as illustrated in Figure \ref{fig:ISE_other_factors}(a, b, c, d) (orange), the DNP enhancement generated by the DQ excitation gets attenuated by the ZQ excitation resulting in a differentiated SE (DSE). 

\begin{figure}[h!]
    \centering
    \includegraphics[width = 0.45\textwidth]{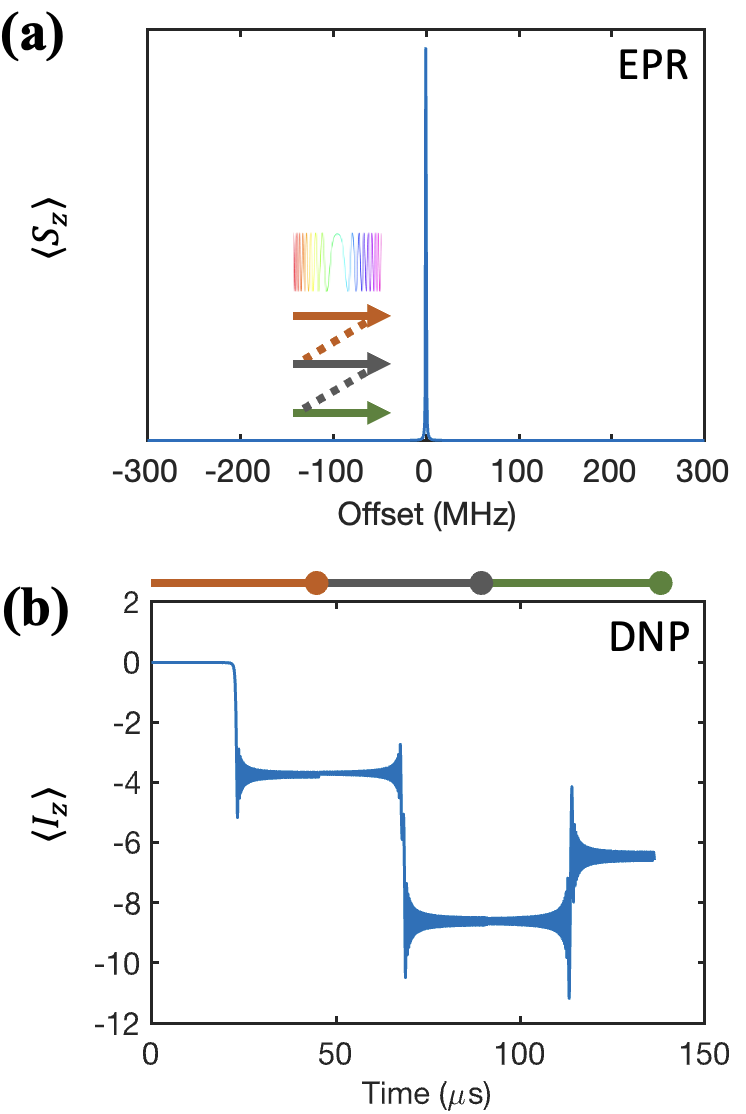}
    \caption{(a) The repeated sweeps through the DQ matching condition is illustrated, to observe the direction of the (b) DNP enhancement build-up due to the coherences evolved in $\rho$ during each sweep. $\omega_{0n} = 100 MHz$, $B = 2.25 MHz$, $\omega_1 = 3.172 MHz$, $k = 2.2 MHz/\mu s$. }
    \label{fig:ISE_cts_sweep}
\end{figure}

We also examined the role of multiple sweeps of the $\mu w$  chirp in ASE experiments where each chirp sweep over a selected DNP transition is expected to integrate the DNP enhancement.\cite{tan2020adiabatic} The assumption was that recurrent DQ transitions would lead to the accumulation of net DNP transfer in the same direction. However, as is seen in Figure \ref{fig:ISE_cts_sweep}(a, b), the DNP enhancement due to the DQ transition was negative in the beginning (brown block), but subsequent chirps resulted in positive (gray block) and then negative (green block) enhancements. The observations presented in Figures 7 and 8 are initially puzzling and appear to contradict the expected outcomes of chirped DNP. However, they can be comprehended if the impact of coherence generated during DQ and ZQ passage is considered.

\subsection{Microscopic Analysis of Coherence Generated During $\mu w$ Chirp Irradiation}

\begin{figure}[h!]
    \centering
    \includegraphics[width = 0.9\textwidth]{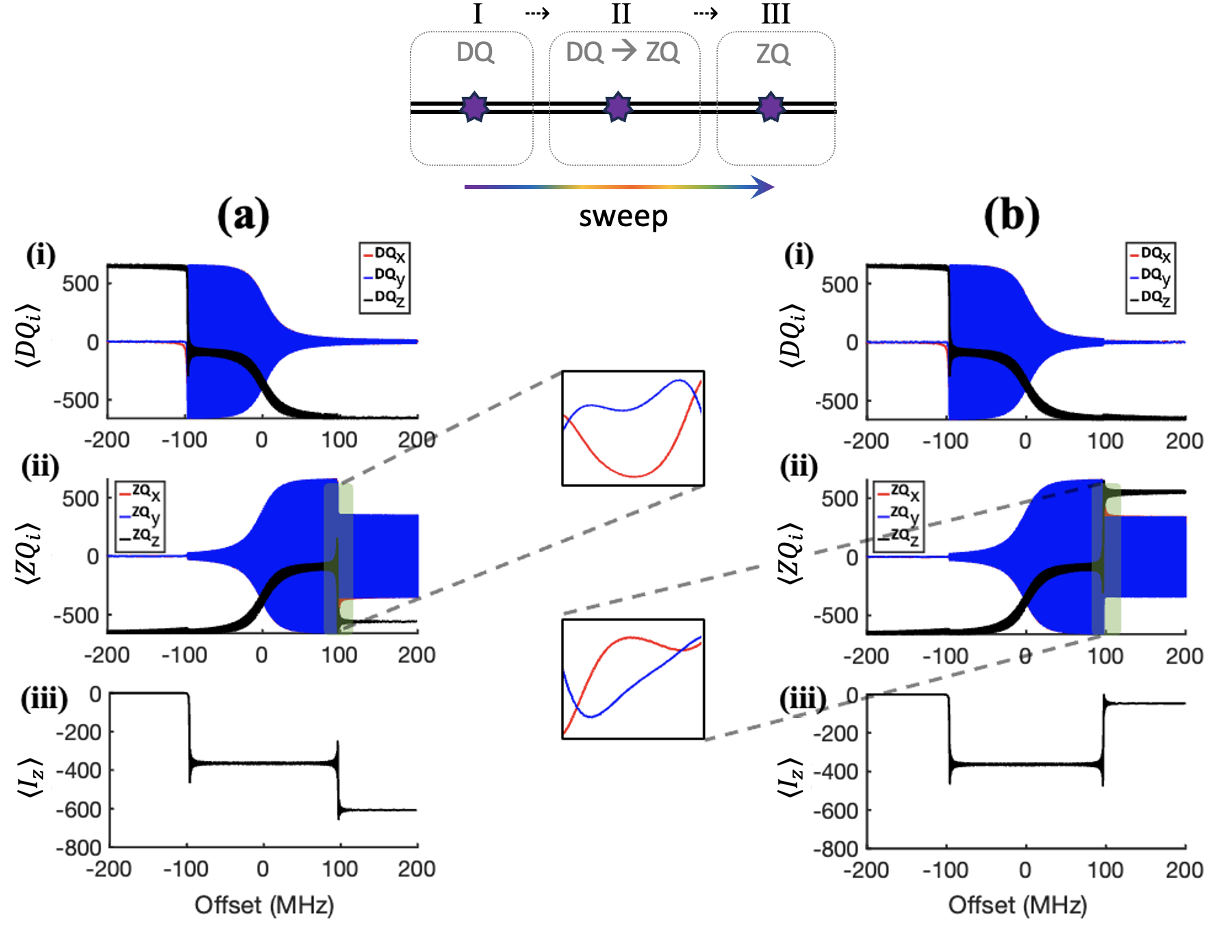}
    \caption{Graphical representation of the evolution of coherences in (i) the DQ subspace, (ii) the ZQ subspace, and (iii) the net nuclear spin (\textbf{I}) magnetization along the z-axis $<I_z>$ during the occurrence of (a). ISE and (b). DSE. $-200 MHz$ to $200 MHz$, $k = 1 MHz/\mu s$, $\omega_{0n} = 100 MHz$, $B = 2.25 MHz$, $\omega_1 (ISE) = 25.15 MHz$, and $\omega_1 (DSE) = 25 MHz$.\\ Pictorial representation of the $\mu w$ chirp pulse partitioned into three regions, (I) $-\delta$ till the DQ transition, (II) after DQ transition till before ZQ transition, and (III) ZQ transition till $\delta$.}
\label{fig:ISE_subspaces}
\end{figure}

To consider the impact of coherences generated while exciting the DQ and ZQ transitions through $\mu w$ chirp irradiation, we divide the density matrix into two subspaces representing the DQ and ZQ transitions as described in Figure \ref{matrix_representation}, Sec. \ref{sec:The Solid Effect}. The time evolution in the independent DQ and ZQ subspaces and the net polarization transferred from \textbf{S} to \textbf{I} for the ISE and DSE scenarios are shown in Figure \ref{fig:ISE_subspaces} (a, b), respectively. 

In order to provide a lucid explanation, we break down the offset axis into three different segments of the $\mu w$ chirp pulse. Assume that the starting offset frequency is $-\delta$ (large negative) and the ending offset frequency is $+\delta$ (large positive); then, the impact of the chirp pulse on the $e-n$ system can be represented within the three offset intervals as shown in Figure \ref{fig:ISE_subspaces}.  

In segment I (illustrated in Figure \ref{fig:ISE_subspaces}), if the swept DQ resonance is adiabatic, the initial density matrix, which is electron spin polarization, gets fully converted into nuclear spin polarization (in the opposite direction) as seen in Figure SI-1. On the Bloch sphere, the magnetization vector in the DQ subspace rotates from $DQ_z$ to $-DQ_z$. Therefore, no coherences are generated. However, extending the chirp offset width to ZQ condition is deleterious as the polarization of nucleus will revert back to electron. Thus, achieving such an adiabatic condition for SE transition is impractical, as it would require either very slow sweep or high $\mu w$ power but electron relaxation and $\mu w$ power are limiting.

In a more practical scenario, the offset sweep across the DQ resonance is non-adiabatic. The density matrix evolution, as shown in Figure \ref{fig:ISE_subspaces} (a(i), b(i)), converts the initial polarization into a combination of polarization ($DQ_z$) and coherences ($DQ_x$ and $DQ_y$), as discussed in Sec. \ref{sec:Adiabatic Inversion}.

Segment II covers the SQ transition of the electron spin \textbf{S} in our $e-n$ system. If the swept SQ resonance is adiabatic, the DQ subspace symmetrically gets converted to the ZQ subspace as $S_z$ adiabatically transforms into $-S_z$ (see Sec. \ref{sec:Adiabatic Inversion}). This is indeed observed in Figure \ref{fig:ISE_subspaces}(a(i, ii), b(i, ii)) where clearly, $DQ_x$, $DQ_y$ and $DQ_z$ swap with $ZQ_x$, $ZQ_y$ and $ZQ_z$.

However, if this transition is non adiabatic, as explained in Sec. \ref{sec:Adiabatic Inversion}, it fails to achieve full inversion of the $S_z$ magnetization. This results in the partial transfer of polarization ($DQ_z$ to $ZQ_z$) and coherences ($DQ_x, DQ_y$ to $ZQ_x, ZQ_y$) from the DQ subspace to the ZQ subspace. This scenario will happen if $\omega_1 \ll k$. It must be noted that it is feasible to adiabatically sweep with respect to the SQ transition.

Consequently, at this point in the chirp, three different scenarios develop that determine the density matrix in the ZQ subspace before the ZQ transition occurs. They are listed in Table 2. 

\begin{table}[h!]
    \centering
    \caption{Tabular representation of the possible outcomes of DQ and SQ transitions in the ZQ subspace. The adiabaticity of the DQ and SQ transitions lead to a density matrix in the ZQ subspace which is either a polarization or a combination of polarization and coherence before the ZQ transition occurs.}
    \begin{tabular}{|c|c|c|c|}
        \hline
        \textbf{Case} & \textbf{DQ Transition} & \textbf{SQ Transition} & \textbf{$\rho^{ZQ}_{DNP}$ Before ZQ Transition} \\
        \hline
        I & Non-Adiabatic & Adiabatic & Polarization + Coherence \\
        \hline
        II & Non-Adiabatic & Non-Adiabatic & Polarization + Coherence \\
        \hline
        III & Adiabatic & Adiabatic & Polarization \\
        \hline
    \end{tabular}
\end{table}

Segment III (ref. Figure \ref{fig:ISE_subspaces}) of the $\mu w$ chirp pulse sweeps through the ZQ SE transition. The outcome of this transition, just like any transition, is determined by: (i) the initial density matrix (in the ZQ subspace) before the transition (given in Table 2), and (ii) the adiabaticity of the transition, as discussed in Sec. \ref{sec:Adiabatic Inversion}. It must be noted that the perturbation driving the ZQ transition is similar to the DQ transition (see Sec. \ref{sec:The Solid Effect}). This implies that, if the DQ transition is adiabatic, the ZQ transition is also adiabatic and vice versa according to the Landau-Zener condition. 

\begin{figure}[h!]
    \centering
    \includegraphics[scale = 0.45]{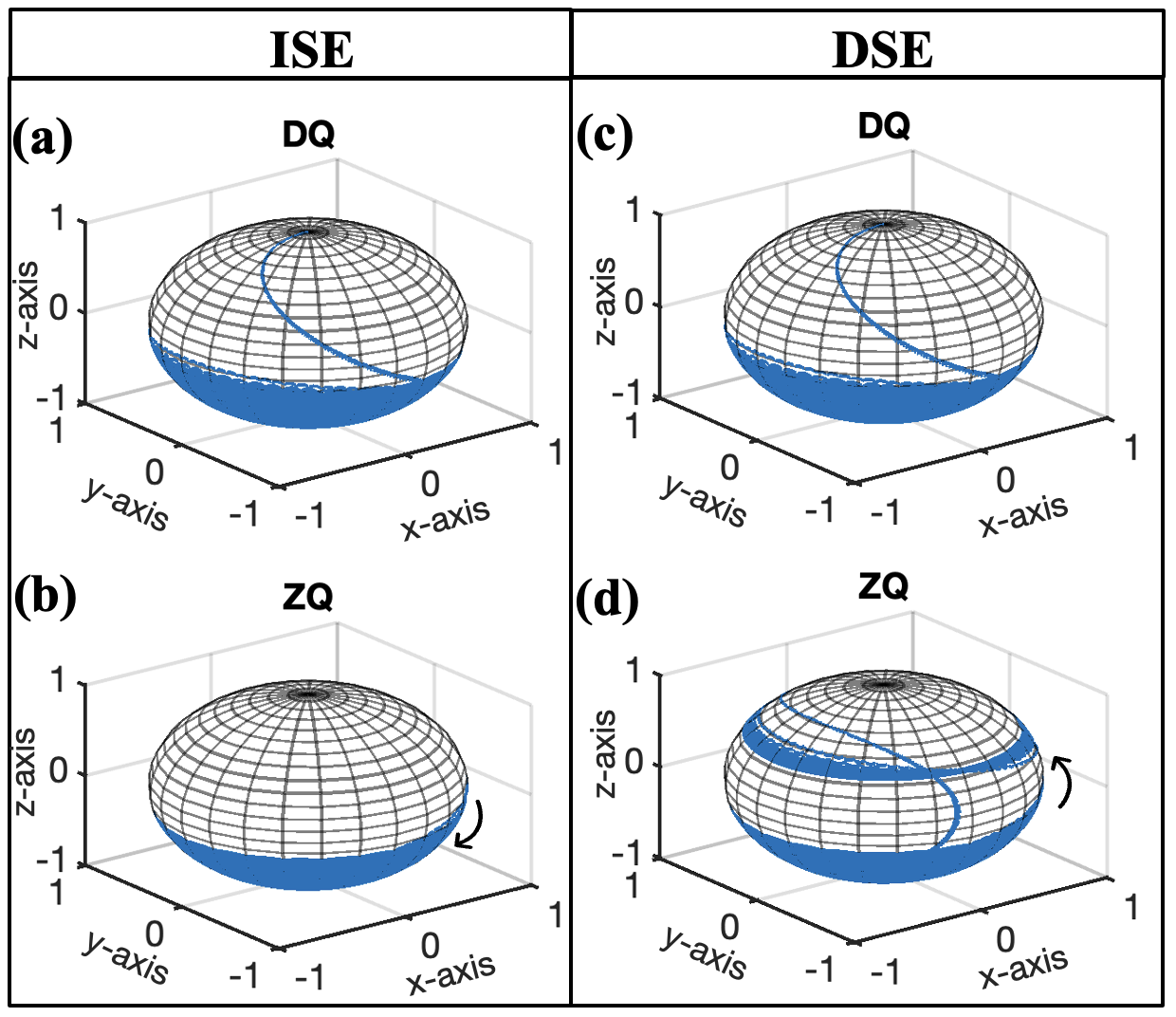}
    \caption{This figure illustrates how the net magnetization vector evolves during ISE in panels (a) for the DQ subspace and (b) for the ZQ subspace, as well as during DSE in panels (c) for the DQ subspace and (d) for the ZQ subspace. Importantly, the magnetization vector in the ZQ subspace can have a net negative z-component leading to ISE, or a net positive z-component leading to DSE. Sweep range: $-200 MHz$ to $200 MHz$, $k = 1 MHz/\mu s$, $\omega_{0n} = 100 MHz$, $B = 2.25 MHz$, $\omega_1 (ISE) = 25.15 MHz$, and $\omega_1 (DSE) = 25 MHz$ }
    \label{fig:bloch_ise_dse}
\end{figure}

As a result, we explore the outcome of the ZQ transition on the three cases discussed in Table 2. For Case I, a non-adiabatic ZQ transition (since the DQ transition is non adiabatic) transforms the initial density matrix in the ZQ subspace, having both polarization ($ZQ_z$) and coherence ($ZQ_x$ and $ZQ_y$), into coherence and positive or negative polarization, analogous to the spin-1/2 case discussed in Sec. \ref{sec:Adiabatic Inversion}. 

The outcome of the ZQ transition on Case II is similar to Case I and also leads to a density matrix (in the ZQ subspace) that comprises of coherence and polarization (positive or negative). The only difference is that the magnitude of polarization transferred is smaller because the non-adiabatic SQ transition leads to the partial transfer of polarization and coherences from the DQ subspace to the ZQ subspace. Case III is an impractical scheme (analyzed in SI).

Therefore, when the DNP transitions (DQ and ZQ) are non-adiabatic, the net transfer of polarization (enhancement) generated by exciting the ZQ transition can either be positive or negative. We observe that: when $ZQ_y$ is greater than $ZQ_x$, net enhancement generated by the ZQ transition is negative and adds up to the DQ enhancement (leading to ISE) (Figure \ref{fig:ISE_subspaces}(a(ii (zoomed in), iii)); otherwise, it is positive and compensates the DQ enhancement (leading to DSE) (Figure \ref{fig:ISE_subspaces}(b(ii (zoomed in), iii)). This is also clearly illustrated in the the evolution of the DQ and ZQ magnetization vectors given in Figure \ref{fig:bloch_ise_dse}. Importantly, the direction of the magnetization vector in the ZQ subspace before the ZQ transition and its rotation due to the transition is responsible for the caveat observed in Figure 7. 

Moreover, the interesting buildup of ASE observed in Figure 8 can also be explained using the same principle. The first sweep through the DQ resonance is non-adiabatic, thus generating a density matrix in the DQ subspace with both polarization and coherence. Subsequent non-adiabatic excitations of the DQ transition generates coherence and a net bi-directional (positive or negative) polarization. Refer to the spin-1/2 scenario explained in Sec. \ref{sec:Adiabatic Inversion}. 

\subsection{Effect of Decoherence} 

Clearly, it is the coherences in the density matrix generated due to the non-adiabatic DQ transition that leads to either the enhancement or the attenuation of nuclear polarization in chirped DNP mechanisms (ISE and ASE). We therefore consider the effect of $T_2$ relaxation, characterized by the loss of phase coherence between spins, on the net polarization transferred from \textbf{S} to \textbf{I} through chirped DNP mechanisms.

\begin{figure}[h!]
    \centering
    \includegraphics[width = \textwidth]
    {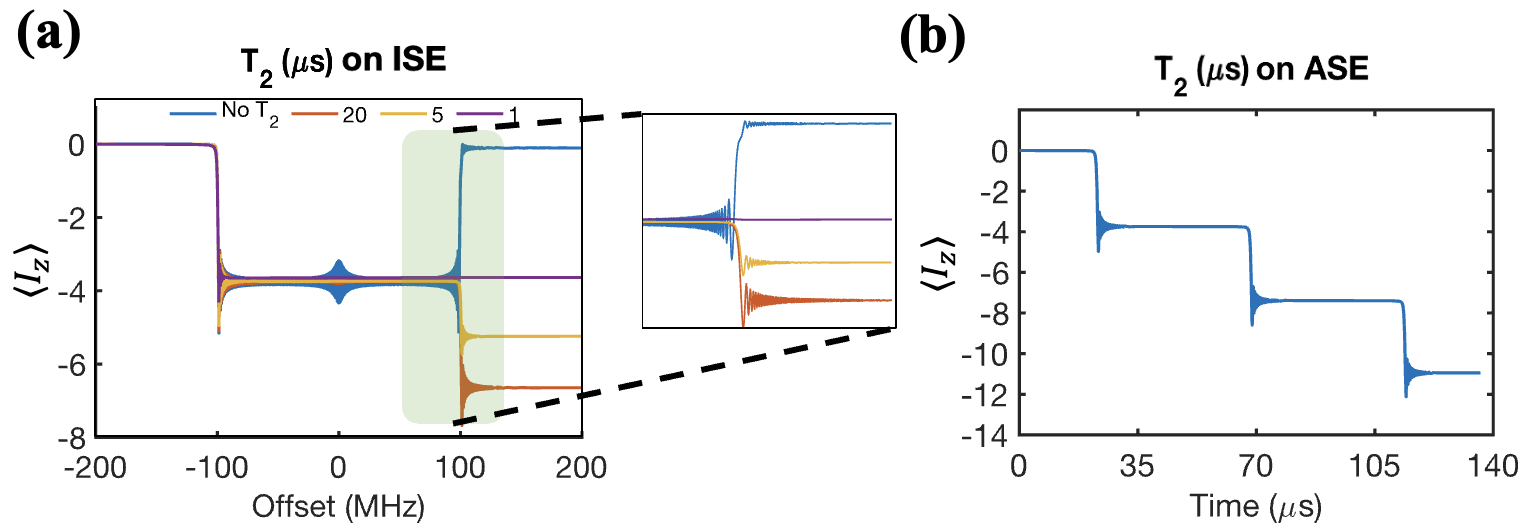}
    \caption{Illustration of (a) consistent occurrence of ISE DNP when the optimal $T_2$ relaxation time is considered, leading to effective integration of DQ and ZQ DNP enhancements. Sweep range: $-200 MHz$ to $200 MHz$, $k = 2.2 MHz/\mu s$, $\omega_{0n} = 100 MHz$, $\omega_1 = 3.172 MHz$, and $B = 2.25 MHz$ (b) Unidirectional build-up of ASE-DNP when $T_2 = 5 \mu s$ as opposed to the bi-directional build-up seen in Figure \ref{fig:ISE_cts_sweep} with the same sweep parameters.
 }
    \label{fig:ISE_with_T2}
\end{figure}

In the case of ISE, incorporating relaxation of ZQ and DQ coherences into the analysis shows that decoherence is crucial for consistently achieving ISE. For $T_2$ relaxation time constants ranging from one to a few hundred $\mu s$, ISE is consistently observed irrespective of factors such as $\omega_1$, $k$, $\omega_{0n}$ and $\beta_{en}$, with a single case shown in Figure \ref{fig:ISE_with_T2}a as an example. Moreover, one also observes a unidirectional build-up of polarization during ASE-DNP under $T_2$ relaxation as seen in Figure \ref{fig:ISE_with_T2}b. This is in contrast with the occasional attenuation of polarization observed in Figure \ref{fig:ISE_cts_sweep}b when $T_2$ relaxation was not considered. 

Hence, it can be inferred that, for ISE, $T_2$ must be optimal; slow enough to allow the DQ transition to occur, yet sufficiently fast to dissipate the coherence prior to the ZQ transition, as illustrated in Figure \ref{fig:ISE_with_T2}a(zoomed in). Likewise, for ASE, $T_2$ relaxation should also be optimum to observe maximum enhancement build-up. This is because very short $T_2$ relaxation times lead to poor DNP enhancement.

\section{Conclusion}
In this study, we have provided a comprehensive analysis of the theory and dynamics underlying chirped Dynamic Nuclear Polarization (DNP) techniques, focusing on the dependence of transition outcomes on their adiabaticity. Although chirped DNP has been widely applied experimentally, several fundamental aspects remain unexplored. We identified a key inconsistency in the the Integrated Solid Effect (ISE), a variant of chirped DNP - its occurrence was being influenced by experimental parameters such as sweep rate, $\mu w$ amplitude, electron-nucleus orientation, and external magnetic field. Our investigation showed that the quantum coherence resulting from double-quantum (DQ) and zero-quantum (ZQ) transitions are causing this observed caveat, with the coherence being triggered by non-adiabatic transitions. Furthermore, we examined the role of optimal decoherence in decaying  quantum coherence, revealing how it facilitates the consistent observation of the ISE. Optimizing the control over coherence decay can lead to more reliable and efficient chirped DNP techniques.

Hence, this work lays the foundation for future studies aimed at improving the consistency and effectiveness of chirped DNP methods, with significant implications for advancing hyperpolarization strategies. It also opens up new possibilities for utilizing room-temperature hyperpolarization in biomedical applications, particularly with photoexcited triplet electrons. The entire analysis can also be applied to magic-angle spinning DNP, where the resonances occur at different times and the overall DNP effect is accumulated over multiple rotor cycles.\cite{thurber2012theory,mentink2015theoretical,equbal2021role}

\subsection*{Acknowledgments}

The authors would like to thank New York University Abu Dhabi (NYUAD) for the financial support of this work and Core Technology Platforms and High Performance Computing facilities of New York University Abu Dhabi for facilitating experimental and theoretical DNP research. Contribution from AE was supported by Tamkeen under the NYU Abu Dhabi Research Institute grant CG008.

\bibliographystyle{unsrt}
\bibliography{ISE.bib}

\begin{thebibliography}{10}

\bibitem{ernst1990principles}
Richard~R Ernst, Geoffrey Bodenhausen, and Alexander Wokaun.
\newblock {\em Principles of nuclear magnetic resonance in one and two dimensions}.
\newblock Oxford university press, 1990.

\bibitem{kay2011nmr}
Lewis~E Kay.
\newblock Nmr studies of protein structure and dynamics.
\newblock {\em Journal of magnetic resonance}, 213(2):477--491, 2011.

\bibitem{hashemi2012mri}
Ray~Hashman Hashemi, William~G Bradley, and Christopher~J Lisanti.
\newblock {\em MRI: the basics: The Basics}.
\newblock Lippincott Williams \& Wilkins, 2012.

\bibitem{qiang2017structural}
Wei Qiang, Wai-Ming Yau, Jun-Xia Lu, John Collinge, and Robert Tycko.
\newblock Structural variation in amyloid-$\beta$ fibrils from alzheimer's disease clinical subtypes.
\newblock {\em Nature}, 541(7636):217--221, 2017.

\bibitem{grey2004nmr}
Clare~P Grey and Nicolas Dupr{\'e}.
\newblock Nmr studies of cathode materials for lithium-ion rechargeable batteries.
\newblock {\em Chemical reviews}, 104(10):4493--4512, 2004.

\bibitem{abragam1961principles}
Anatole Abragam.
\newblock {\em The principles of nuclear magnetism}.
\newblock Number~32. Oxford university press, 1961.

\bibitem{abragam1978principles}
Anatole Abragam and Maurice Goldman.
\newblock Principles of dynamic nuclear polarisation.
\newblock {\em Reports on Progress in Physics}, 41(3):395, 1978.

\bibitem{eills2023spin}
James Eills, Dmitry Budker, Silvia Cavagnero, Eduard~Y Chekmenev, Stuart~J Elliott, Sami Jannin, Anne Lesage, Jorg Matysik, Thomas Meersmann, Thomas Prisner, et~al.
\newblock Spin hyperpolarization in modern magnetic resonance.
\newblock {\em Chemical reviews}, 123(4):1417--1551, 2023.

\bibitem{maly2008dynamic}
Thorsten Maly, Galia~T Debelouchina, Vikram~S Bajaj, Kan-Nian Hu, Chan-Gyu Joo, Melody~L Mak-Jurkauskas, Jagadishwar~R Sirigiri, Patrick~CA Van Der~Wel, Judith Herzfeld, Richard~J Temkin, et~al.
\newblock Dynamic nuclear polarization at high magnetic fields.
\newblock {\em The Journal of chemical physics}, 128(5), 2008.

\bibitem{barnes2008high}
AB~Barnes, G~De~Pa{\"e}pe, PCA Van Der~Wel, K-N Hu, C-G Joo, VS~Bajaj, ML~Mak-Jurkauskas, JR~Sirigiri, J~Herzfeld, RJ~Temkin, et~al.
\newblock High-field dynamic nuclear polarization for solid and solution biological nmr.
\newblock {\em Applied magnetic resonance}, 34:237--263, 2008.

\bibitem{carver1953polarization}
Tom~R Carver and Charles~P Slichter.
\newblock Polarization of nuclear spins in metals.
\newblock {\em Physical Review}, 92(1):212, 1953.

\bibitem{overhauser1953polarization}
Albert~W Overhauser.
\newblock Polarization of nuclei in metals.
\newblock {\em Physical Review}, 92(2):411, 1953.

\bibitem{hovav2010theoretical}
Yonatan Hovav, Akiva Feintuch, and Shimon Vega.
\newblock Theoretical aspects of dynamic nuclear polarization in the solid state--the solid effect.
\newblock {\em Journal of Magnetic Resonance}, 207(2):176--189, 2010.

\bibitem{hovav2012theoretical}
Yonatan Hovav, Akiva Feintuch, and Shimon Vega.
\newblock Theoretical aspects of dynamic nuclear polarization in the solid state--the cross effect.
\newblock {\em Journal of Magnetic Resonance}, 214:29--41, 2012.

\bibitem{thankamony2017dynamic}
Aany Sofia~Lilly Thankamony, Johannes~J Wittmann, Monu Kaushik, and Bj{\"o}rn Corzilius.
\newblock Dynamic nuclear polarization for sensitivity enhancement in modern solid-state nmr.
\newblock {\em Progress in nuclear magnetic resonance spectroscopy}, 102:120--195, 2017.

\bibitem{hu2011quantum}
Kan-Nian Hu, Galia~T Debelouchina, Albert~A Smith, and Robert~G Griffin.
\newblock Quantum mechanical theory of dynamic nuclear polarization in solid dielectrics.
\newblock {\em The Journal of chemical physics}, 134(12), 2011.

\bibitem{corzilius2012solid}
Bj{\"o}rn Corzilius, Albert~A Smith, and Robert~G Griffin.
\newblock Solid effect in magic angle spinning dynamic nuclear polarization.
\newblock {\em The Journal of chemical physics}, 137(5), 2012.

\bibitem{takeda2001dynamic}
Kazuyuki Takeda, K~Takegoshi, and Takehiko Terao.
\newblock Dynamic nuclear polarization by photoexcited-triplet electron spins in polycrystalline samples.
\newblock {\em Chemical physics letters}, 345(1-2):166--170, 2001.

\bibitem{hamachi2024recent}
Tomoyuki Hamachi and Nobuhiro Yanai.
\newblock Recent developments in materials and applications of triplet dynamic nuclear polarization.
\newblock {\em Progress in Nuclear Magnetic Resonance Spectroscopy}, 2024.

\bibitem{wenckebach2008solid}
W~Th Wenckebach.
\newblock The solid effect.
\newblock {\em Applied Magnetic Resonance}, 34(3):227--235, 2008.

\bibitem{kaminker2018amplification}
Ilia Kaminker and Songi Han.
\newblock Amplification of dynamic nuclear polarization at 200 ghz by arbitrary pulse shaping of the electron spin saturation profile.
\newblock {\em The journal of physical chemistry letters}, 9(11):3110--3115, 2018.

\bibitem{equbal2019pulse}
Asif Equbal, Kan Tagami, and Songi Han.
\newblock Pulse-shaped dynamic nuclear polarization under magic-angle spinning.
\newblock {\em The Journal of Physical Chemistry Letters}, 10(24):7781--7788, 2019.

\bibitem{can2018frequency}
TV~Can, JE~McKay, RT~Weber, C~Yang, T~Dubroca, J~Van~Tol, S~Hill, and RG~Griffin.
\newblock Frequency-swept integrated and stretched solid effect dynamic nuclear polarization.
\newblock {\em The journal of physical chemistry letters}, 9(12):3187--3192, 2018.

\bibitem{quan2022integrated}
Yifan Quan, Jakob Steiner, Yifu Ouyang, Kong~Ooi Tan, W~Thomas Wenckebach, Patrick Hautle, and Robert~G Griffin.
\newblock Integrated, stretched, and adiabatic solid effects.
\newblock {\em The journal of physical chemistry letters}, 13(25):5751--5757, 2022.

\bibitem{henstra1988enhanced}
A~Henstra, P~Dirksen, and W~Th Wenckebach.
\newblock Enhanced dynamic nuclear polarization by the integrated solid effect.
\newblock {\em Physics Letters A}, 134(2):134--136, 1988.

\bibitem{HENSTRA19906}
A.~Henstra, T.-S. Lin, J.~Schmidt, and W.Th. Wenckebach.
\newblock High dynamic nuclear polarization at room temperature.
\newblock {\em Chemical Physics Letters}, 165(1):6--10, 1990.

\bibitem{henstra2014dynamic}
A~Henstra and W~Th Wenckebach.
\newblock Dynamic nuclear polarisation via the integrated solid effect i: theory.
\newblock {\em Molecular Physics}, 112(13):1761--1772, 2014.

\bibitem{iinuma2000high}
Masataka Iinuma, Yoshiro Takahashi, Ippei Shake, Masahiro Oda, Akira Masaike, Tsutomu Yabuzaki, and Hirohiko~M Shimizu.
\newblock High proton polarization by microwave-induced optical nuclear polarization at 77 k.
\newblock {\em Physical Review Letters}, 84(1):171, 2000.

\bibitem{nishimura2020materials}
Koki Nishimura, Hironori Kouno, Yusuke Kawashima, Kana Orihashi, Saiya Fujiwara, Kenichiro Tateishi, Tomohiro Uesaka, Nobuo Kimizuka, and Nobuhiro Yanai.
\newblock Materials chemistry of triplet dynamic nuclear polarization.
\newblock {\em Chemical Communications}, 56(53):7217--7232, 2020.

\bibitem{tan2020adiabatic}
Kong~Ooi Tan, Ralph~T Weber, Thach~V Can, and Robert~G Griffin.
\newblock Adiabatic solid effect.
\newblock {\em The journal of physical chemistry letters}, 11(9):3416--3421, 2020.

\bibitem{quan2023general}
Yifan Quan, Nemanja Niketic, Jakob~M Steiner, Tim~R Eichhorn, W~Tom~Wenckebach, and Patrick Hautle.
\newblock General theory of light propagation and triplet generation for studies of spin dynamics and triplet dynamic nuclear polarisation.
\newblock {\em Molecular Physics}, 121(3):e2169025, 2023.

\bibitem{hautle2024creating}
W~Th Wenckebach.
\newblock Creating high, portable proton polarization with photo-excited triplet dnp.
\newblock {\em Journal of Magnetic Resonance Open}, page 100159, 2024.

\bibitem{jain2017off}
Sheetal~K Jain, Guinevere Mathies, and Robert~G Griffin.
\newblock Off-resonance novel.
\newblock {\em The Journal of Chemical Physics}, 147(16), 2017.

\bibitem{pang2022unified}
Zhenfeng Pang, Sheetal Jain, Chen Yang, Xueqian Kong, and Kong~Ooi Tan.
\newblock A unified description for polarization-transfer mechanisms in magnetic resonance in static solids: Cross polarization and dnp.
\newblock {\em The Journal of Chemical Physics}, 156(24), 2022.

\bibitem{vega1978fictitious}
Shimon Vega.
\newblock Fictitious spin 1/2 operator formalism for multiple quantum nmr.
\newblock {\em The Journal of Chemical Physics}, 68(12):5518--5527, 1978.

\bibitem{ivanov2021floquet}
Konstantin~L Ivanov, Kaustubh~R Mote, Matthias Ernst, Asif Equbal, and Perunthiruthy~K Madhu.
\newblock Floquet theory in magnetic resonance: Formalism and applications.
\newblock {\em Progress in Nuclear Magnetic Resonance Spectroscopy}, 126:17--58, 2021.

\bibitem{baum1985broadband}
J~Baum, R~Tycko, and ANDA Pines.
\newblock Broadband and adiabatic inversion of a two-level system by phase-modulated pulses.
\newblock {\em Physical Review A}, 32(6):3435, 1985.

\bibitem{garwood2001return}
Michael Garwood and Lance DelaBarre.
\newblock The return of the frequency sweep: designing adiabatic pulses for contemporary nmr.
\newblock {\em Journal of magnetic resonance}, 153(2):155--177, 2001.

\bibitem{thurber2012theory}
Kent~R Thurber and Robert Tycko.
\newblock Theory for cross effect dynamic nuclear polarization under magic-angle spinning in solid state nuclear magnetic resonance: The importance of level crossings.
\newblock {\em The Journal of chemical physics}, 137(8), 2012.

\bibitem{mentink2015theoretical}
Frederic Mentink-Vigier, Uemit Akbey, Hartmut Oschkinat, Shimon Vega, and Akiva Feintuch.
\newblock Theoretical aspects of magic angle spinning-dynamic nuclear polarization.
\newblock {\em Journal of Magnetic Resonance}, 258:102--120, 2015.

\bibitem{equbal2021role}
Asif Equbal, Sheetal~Kumar Jain, Yuanxin Li, Kan Tagami, Xiaoling Wang, and Songi Han.
\newblock Role of electron spin dynamics and coupling network in designing dynamic nuclear polarization.
\newblock {\em Progress in Nuclear Magnetic Resonance Spectroscopy}, 126:1--16, 2021.

\end{thebibliography}

\end{document}